\documentclass[sigplan,nonacm]{acmart}

\usepackage{amsmath}
\usepackage[inline]{enumitem}   
\usepackage[normalem]{ulem}  
\usepackage{xurl}  
\usepackage{graphicx}
\usepackage{subcaption}
\usepackage[frozencache,cachedir=minted-cache]{minted}
\usepackage{makecell}  
\usepackage{multirow}  
\usepackage{diagbox}   
\usepackage{booktabs}
\usepackage{wrapfig}

\setlist[itemize]{leftmargin=1em, itemsep=0pt, topsep=2pt plus 2pt, parsep=2pt}
\setlist[enumerate]{leftmargin=1em, itemsep=0pt, topsep=2pt plus 2pt, parsep=2pt}

\definecolor{mauve}{HTML}{9558B2}
\definecolor{rose}{HTML}{FF007F}

\newcommand{\XComment}[1]{}
\ifdefined\COMMENTSON
\newcommand{\yifan}[1]{\textcolor{rose}{Yifan: #1}}
\newcommand{\sasa}[1]{\textcolor{mauve}{#1}}
\newcommand{\yuchen}[1]{\textcolor{orange}{Yuchen: #1}}
\newcommand{\matei}[1]{\textcolor{purple}{Matei: #1}}
\newcommand{\egan}[1]{{\color{green} Egan: #1}}
\newcommand{\minsoo}[1]{{\color{red} Minsoo: #1}}

\newcommand{\NA}[1]{{\color{blue} #1}}
\else
\newcommand{\yifan}[1]{}
\newcommand{\sasa}[1]{}
\newcommand{\yuchen}[1]{}
\newcommand{\matei}[1]{}
\newcommand{\egan}[1]{}
\newcommand{\minsoo}[1]{}

\newcommand{\NA}[1]{#1}
\fi

\def\NAME{Nautilus}

\begin{document}

\date{}
\title{Nautilus: An Auto-Scheduling Tensor Compiler for Efficient Tiled GPU Kernels}

\author{Yifan Zhao}
\affiliation{
  \institution{University of Illinois Urbana-Champaign}
  \country{USA}
}
\email{yifanz16@illinois.edu}

\author{Yuchen Yang}
\affiliation{
  \institution{University of Illinois Urbana-Champaign}
  \country{USA}
}
\email{yucheny8@illinois.edu}

\author{Matei Budiu}
\affiliation{
  \institution{University of Illinois Urbana-Champaign}
  \country{USA}
}
\email{mbudiu2@illinois.edu}

\author{Sasa Misailovic}
\affiliation{
  \institution{University of Illinois Urbana-Champaign}
  \country{USA}
}
\email{misailo@illinois.edu}

\begin{abstract}
  We present \NAME{}, a novel tensor compiler that moves toward fully automated
  \emph{math-to-kernel} optimization.
  \NAME{} compiles a high-level algebraic specification of tensor operators into efficient tiled GPU kernels.
  \NAME{}'s successive lowering design allows high-level optimizations, expression
  rewrites, and tile optimizations to be jointly applied in a single end-to-end system.
  \NAME{} presents a novel auto-scheduler that discovers sequences of high-level
  optimizations, while preserving the regular program structure
  needed by tile optimizers.
  \NAME{}'s auto-scheduler captures complex interactions and trade-offs in the high-level
  optimizations,
  including aggressive global transformations like advanced reduction fusion.

  \NAME{} is the first end-to-end tensor compiler capable of starting from a math-like
  description of attention
  and automatically discovering FlashAttention-3-like kernels,
  offloading the entire burden of optimization from the programmer to the compiler.
  Across five transformer-based models and 150 evaluation configurations on NVIDIA GH200 and RTX
  5090 GPUs, \NAME{} achieves up to 23\% higher throughput than state-of-the-art compilers on GH200
  and up to 42\% on RTX 5090, while matching or exceeding manually written cuDNN kernels
  on many \mbox{long-sequence configurations.}
\end{abstract}

\maketitle

\section{Introduction}
{Modern attention-based machine learning (ML) models
are essential for language, vision, and other emerging tasks in artificial intelligence
\cite{gpt5,image-gen-paper,audio-gen-paper}.
To efficiently deploy these models on GPUs, system engineers need tools to bridge the performance gap
between a \emph{math-like model specification}, such as PyTorch or JAX code, and efficient
low-level code that maximally utilizes the underlying hardware.
Today, the connection between math-like specification and efficient execution on increasingly varied hardware
is challenged ---
existing tools are either too hard to use, requiring expertise in rapidly-evolving low-level
hardware details, or produce \mbox{programs that are not sufficiently performant.}

The most common approach today is writing manually optimized kernels in hardware vendor languages
like CUDA~\cite{cuda} and CUTLASS~\cite{cutlass}. They achieve high performance and have been
used in PyTorch~\cite{pytorch} and TensorFlow~\cite{tensorflow}.
However, programming in these low-level languages requires significant hardware and
algorithmic expertise and does not scale to rapidly evolving GPU architectures and ML models.
On the opposite side, schedule-based tensor compilers such as Halide~\cite{halide} and TVM~\cite{tvm}
optimize a high-level specification of the ML operator {in a math-like tensor language.}
This design offloads the obligation of optimization from the programmer to the compiler,
which needs to search in a complex design space of \emph{schedules}
(sequence of transformations) to find an optimal kernel on the target GPU.
But, an unconstrained, exhaustive search in this design space is intractable, thus
today's high-level tensor compilers fail to match the performance \mbox{of handwritten code}.

As a tentative trade-off between programmability and efficiency,
tile-based compilers and languages,
including Triton~\cite{triton}, Pallas~\cite{pallas}, and TileLang~\cite{tilelang},
have emerged as an abstraction layer over hardware vendor languages.
Here, a programmer manually expresses the ML operator as computation over tiles (fixed size subtensors),
and the tile compiler applies low-level SIMD
optimizations to map tile computation to the hardware.
With tile languages, a programmer still has to manually control many essential
high-level optimizations, making it difficult to write and port to new hardware.
This issue is especially evident for attention, where multiple versions of FlashAttention
\cite{flash-attention,flash-attention-2,flash-decoding,flash-attention-3}
require high-level optimizations that tile compilers cannot automate.
Programmers then need to manually write increasingly more complex tile programs to implement
\NA{those kernels and significantly modify them for every new GPU architecture.}


\XComment{Matei: [this can be maybe moved around a bit] We believe that tile languages do half of
  the job well - simplifying the lower level of kernel generation - and in this paper, we
  finish the job by providing the other half of the work:
  a complier which automates generation of performant tile language code from a
  mathematical description,
  allowing developers for the first time to attain state-of-the-art performance without
  having to worry
  about rapidly changing low-level hardware at all. This is possible for the first time
  thanks to advanced loop
  fusion techniques presented in Neptune, allowing for the automatic generation of
FlashAttention-like fused kernels.}

We advocate that a math-to-kernel compiler \NA{is a viable solution that can yield ML kernel efficiency competitive with manually-written code on modern GPU architectures.}
The programmer should be able to write operators in a math-like linear algebraic language
(e.g., attention as a sequence of matmuls and softmax),
and the compiler automatically finds high-performance kernels (e.g., FlashAttention).
The central difficulty to materialize this goal is that we must bridge two types of optimizations
that live on different levels of abstraction:
(1)~operation-level optimizations that reason about compute expressions and data dependencies of linear algebra operations,
and (2)~tile optimizations that map regular computation patterns to the underlying hardware.

While tile compilers are effective low-level optimizers,
they are suboptimal programming interfaces as they fail to automate operation-level optimizations.
First, compiling many tensor operations into one or a few efficient kernels
requires new fusion techniques to handle complex data dependency.
In particular, \emph{advanced reduction fusion}~\cite{neptune} can fuse reduction loop nests,
such as softmax and matmul in attention,
and it is a key optimization to allow compilers to automatically discover
FlashAttention from math-like attention.
Yet it is still unclear how to apply such optimizations in a compiler without user intervention.
Second, a rigorous translation pipeline and stable intermediate representations (IRs)
are needed to bring the program from math expressions to a kernel, while facilitating optimizations \mbox{at
different abstraction levels.}

\vspace{0.03in}
\noindent{\textbf{Our work: \NAME{}}. } \NAME{} is a novel tensor-compiler for
attention-based ML models.
\NAME{} brings together a variety of optimizations: operation-level optimizations
(advanced reduction fusion),
expression rewrites, and tile optimizations in one pipeline, to generate high-performance
GPU kernels on modern and emerging GPUs.
\NA{These categories of optimizations live on very different abstraction levels.
This work is the first to identify these three categories
as critical stages in end-to-end tensor compilation,
and provide a successively lowering pipeline where each IR 
is at the right abstraction level to facilitate a \mbox{category of optimizations.}}

\NA{Operation-level optimizations, such as advanced reduction fusion,
have been previously challenging to automate without user intervention.
They often perform global, wide-reaching transformations on the program structure,
making it challenging to capture interactions between them,
and they present non-trivial tradeoffs for which there often are no effective heuristics.}
\NAME{} adopts the concept of schedules to regulate the legality of the sequence of transformations
and facilitate search-based exploration~\cite{tvm}.
\NAME{} automates the generation of schedules that incorporate advanced reduction fusion
via a novel auto-scheduler designed specifically to target tile-based languages.
This auto-scheduler maintains regularity in the program structure
to make the program ``tile-ready'', i.e., easy to optimize by tile optimizers. \NA{As a result of our design, tile-based languages naturally land as low-level intermediate \mbox{representations and backends in Nautilus.}}

We demonstrate that \NAME{} can find schedules that automatically discover variants
of FlashAttention3,
starting from a simple linear algebraic representation of ordinary attention operator.
\NAME{} is the first end-to-end compiler to perform fully automated math-to-kernel optimization
on attention to deliver FlashAttention3 performance. 

We evaluate \NAME{} on five
transformer-based models, with {150} total evaluation configurations, on two
modern GPU architectures: NVIDIA GH200 (Hopper) and NVIDIA RTX 5090 (Blackwell). We used
seven compiler backends and five hand-optimized libraries as baselines. \NAME{}-generated attention kernels have up to 23\% higher throughput over the
state-of-the-art compiler systems on NVIDIA GH200 and up to 42\% on NVIDIA RTX 5090.
Moreover, our results are comparable to and for many configurations (with longer sequences)
faster than manually written cuDNN kernels.

\vspace{.03in}
\noindent\textbf{Contributions:} The paper makes \mbox{several contributions:}
\begin{itemize}
  \item We present \NAME{}, a tensor compiler that fully automates GPU kernel generation
    from program description,
    and produces high-performance GPU kernels.
  \item We present an automatic scheduler that discovers optimization schedules using our
    language with a library of
    fusion strategies and other high-level optimizations, from the computation pattern of
    the operator.
  \item We present a three-IR (intermediate representation) abstraction to facilitate
    the wide-ranging optimizations of \NAME{}: a scalar IR and two flavors of tile IRs.
    One tile IR (VR-tile IR) is novel and facilitates expression rewriting optimizations,
    and the other tile IR (MA-tile IR) enables the use of multiple tile-based optimizers in \NAME{}.
  \item We implement \NAME{} and evaluate it on multiple state-of-the-art attention
    models and GPU architectures. Our results show better throughput and latency than
    existing compilers and on par with hand-optimized kernels.
\end{itemize}
}

\section{\NAME{} System Design}
\label{sec:design}
{Figure \ref{fig:design} shows the architecture of \NAME{}.
\NAME{} automatically finds the best schedule for the input program,
applies multiple sets of optimizations to the program, and generates runnable tensor kernels.
\NAME{} has three main components: 
\begin{itemize}
  \item A scheduling language over the \emph{block graph} of a given program
    (\S\ref{sec:block-graph}).
    \NA{This language defines ways to refer to elements on the block graph and to transform them.}
    Scheduling on the block graph captures high-level optimizations such as operator fusion,
    and leaves other optimizations to later stages of \NAME{}'s pipeline.
  \item An auto-scheduler that finds high-performance schedules for an input program on
    given hardware without user intervention (\S\ref{sec:auto-sched}).
    It incrementally builds schedules over multiple \textit{scheduling passes},
    each examining the program to add transform steps to the schedule.
    \NA{The auto-scheduler is designed to create \textit{tile-ready} programs
    that are amenable to being translated and optimized in \NAME{}'s tile IRs.}
  \item A hierarchical optimization pipeline that successively lowers the program over multiple IRs
    and applies a set of optimizations to each IR (\S\ref{sec:opt-pipeline}).
    After auto-scheduling on the scalar IR, \NAME{} lowers the program
    to a virtual register tile IR (VR-tile IR) for expression rewriting,
    and a memory access tile IR (MA-tile IR) for tile-level SIMD optimizations.
    \NAME{}'s scalar IR and MA-tile IR are adapted from Neptune~\cite{neptune},
    while the VR-tile IR is a \NAME{} creation.
    \NAME{} adapts multiple existing tile-based compilers and optimizers
    for tile-level optimizations.
\end{itemize}

\vspace{-.0in}
\noindent{}The input to \NAME{} is a program written in tensor expressions,
concisely describing the tensor operation to be compiled and optimized.
Users can write tensor expressions directly in TVM's tensor expression (TE) language;
alternatively, \NAME{} uses TVM Relax~\cite{relax} to convert DNN layer definitions from
popular deep learning frameworks (e.g., PyTorch, TensorFlow) to tensor expressions.
Other tensor expression languages, e.g., JAX's jaxpr \mbox{language, could be used.}

\begin{figure}[t]
  \centering\vspace{-.03in}
  \includegraphics[width=0.8\columnwidth, trim=0mm 61mm 2mm 0mm, clip,
  page=1]{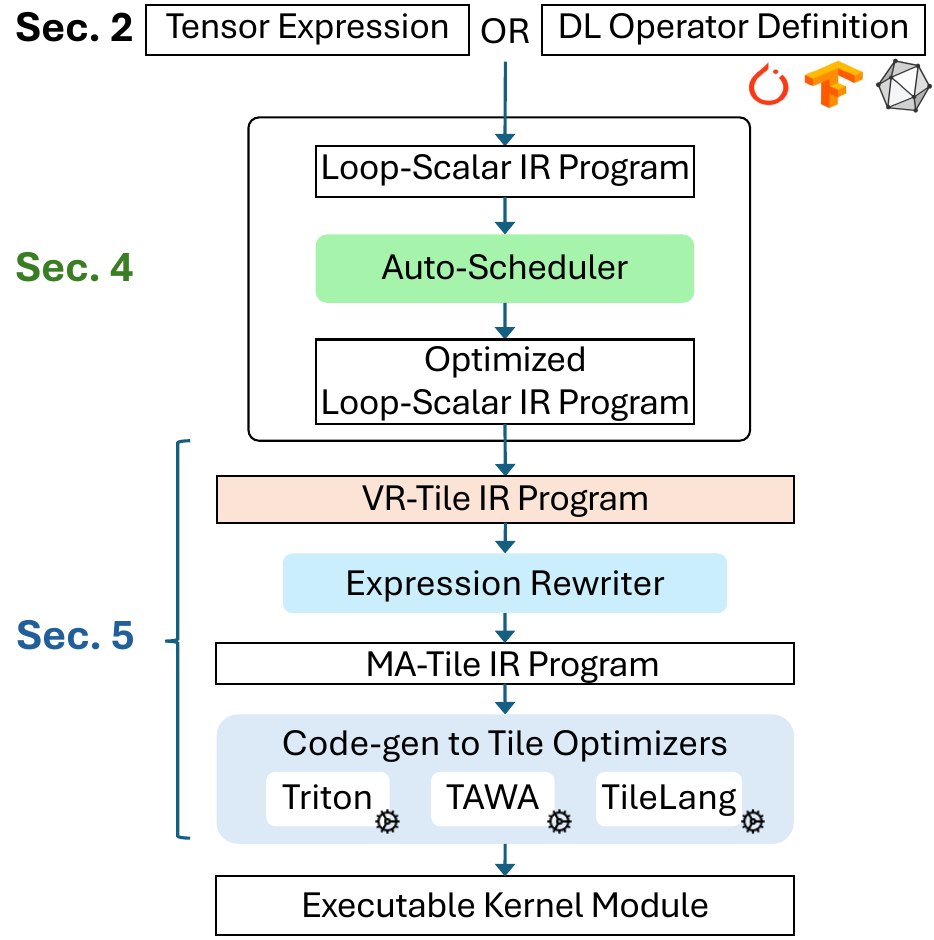}
  \vspace{-.05in}
  \caption{System overview of \NAME{}. Novel components from our contributions are shown in colored boxes.}
  \label{fig:design}\vspace{-.15in}
\end{figure}

\XComment{
\textbf{Input: Tensor Expression.}
\NAME{}'s input programs are written in a compact tensor expression language which
translates straightforwardly into \NAME{}'s scalar IR.
Users can write tensor expressions directly;
alternatively, \NAME{} uses TVM Relax~\cite{relax} to convert DNN layer definitions from
popular deep learning frameworks (e.g., PyTorch, TensorFlow) to tensor expressions.
Tensor expressions are written in TVM's tensor expression (TE) language;
other tensor expression languages, such as JAX's jaxpr language, could be used.
}

\textbf{Block Graph.}
\label{sec:block-graph}
\begin{figure}[t]
  \centering
  \includegraphics[width=0.9\columnwidth, trim=0mm 5mm 25mm 0mm, clip,
  page=2]{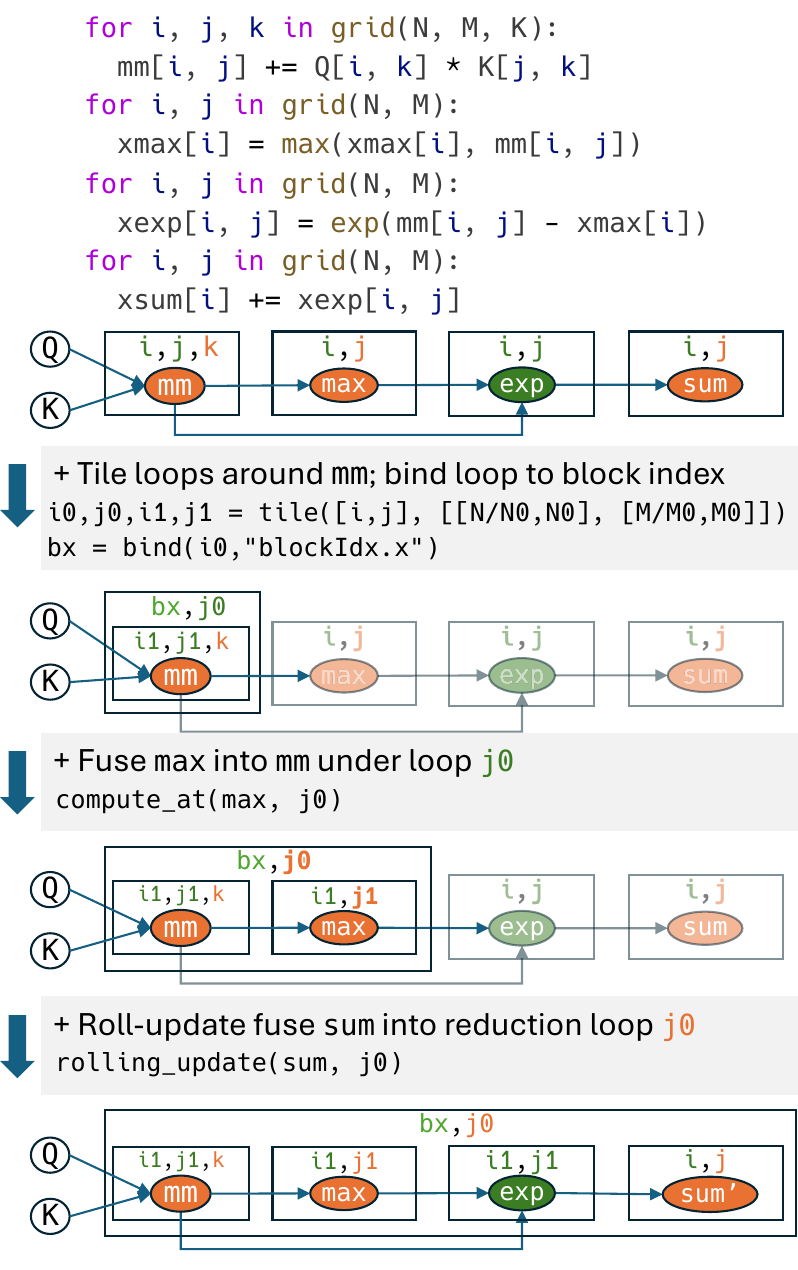}
  \vspace{-.1in}
  \caption{An example of scheduling a block graph.
    A scalar IR program and its corresponding block graph are shown at the top,
    followed by examples of using scheduling primitives to transform the block graph. In
  this graph, ovals represent \emph{compute statements} and rectangles represent \emph{loops}. }
  \label{fig:sched-example}\vspace{-.1in}
\end{figure}
%
\NAME{}'s scheduling focuses on high-level optimizations
and delegates other optimizations to later stages of the pipeline.
A key data structure that captures this separation of optimization is the \emph{block graph}.
The block graph describes the work executed by the entire program and each GPU thread block.
All operations in the scheduling language of \NAME{} are transformations on the block graph.

The block graph combines the data dependency graph and the AST of a scalar IR program.
The top part of Figure \ref{fig:sched-example} shows an example scalar IR program
(extracted from the attention operator) and its block graph.
Each node corresponds to a \emph{compute statement} in the IR,
and directed edges indicate data dependencies.
Boxes around compute statements represent \emph{loops}.
The block graph tracks iteration-carried data dependency of each loop,
determined by the data access pattern of compute statements.
For example, the \texttt{k} loop around the \texttt{mm} node is a reduction loop,
while the loops \texttt{i} and \texttt{j} are map loops.
The block graph also tracks tensors' placements in memory and which loops are mapped to GPU blocks.

The \NAME{} auto-scheduler operates over compute statements and loops.
Additionally, it uses the concepts of \emph{producers}, \emph{predecessors}, and \emph{paths}, which
we define as follows:
The \emph{producers} of a compute statement $n$ are all the other compute statements
which are in-neighbors
of that statement in the block graph (i.e. all compute statements with a directed edge to $n$).
A compute statement $n_2$ is a \emph{predecessor} of compute statement $n_1$
if $n_2$ is
a producer of $n_1$, or if $n_2$ is a producer of any other predecessor of $n_1$.
A \emph{path} between two compute statements $n_1$ and $n_2$ is a set of statements that
are producers of each other which \mbox{begins at $n_1$ and ends at $n_2$.}

}

\vspace{-.05in}
\section{Background}
\label{sec:background}
{\textbf{Schedule-based Tensor Optimizations.}
Schedule-based tensor compilers (e.g., TVM~\cite{tvm} and Halide~\cite{halide})
separate the mathematical definition of a tensor program from \mbox{optimization} decisions.
Optimization decisions are conveyed via dedicated scheduling languages,
with a library of program transformations, e.g., for loops,
parallelization, and data movement.
%
Auto-schedulers (e.g., Ansor~\cite{ansor} and Halide's auto-sche\-duler~\cite{halide-autosched})
aim to relieve users from hand-writing schedules by searching over the space of
legal transformation sequences.
They rely on heuristic rules and cost models to prune unfavorable schedules.
The design of schedule languages adds to the difficulty of auto-scheduling:
they tend to expose a set of fine-grained transformations
and mix optimizations from different levels (e.g., high-level fusion decisions with
low-level thread mappings) \mbox{in a single schedule.}

\textbf{Tile Optimizations.}
Tile-based compilers (e.g., Triton~\cite{triton} and TileLang~\cite{tilelang})
are an alternative approach to effectively automate low-level optimizations.
Programmers describe the workload as tile computation (fixed-shape subtensors)
in a tile-based language with an SIMD-like programming model,
and the compiler maps the tiles to hardware SIMD units.
While tile compilers automate SIMD-level optimizations,
they leave most higher-level optimizations, e.g., operator fusion and expression rewriting,
to the programmer.

\textbf{Bridging Schedule and Tile Optimizations.}
Prior work (e.g., Neptune~\cite{neptune}) has shown that schedule-based and tile-based
approaches are complementary and can combine to achieve higher performance than either.
Schedule languages are well-suited for high-level optimizations such as operator fusion,
and tile compilers excel at low-level mapping of tile programs to hardware.
Neptune proposes a unified framework where scheduling acts on loop-based programs
(loop-scalar IR in Neptune), which Neptune translates into a tile-based representation (tile IR)
to exploit Triton for SIMD optimizations.
Neptune still requires major manual effort to use, as it does not include an auto-scheduler
and requires the user to provide a schedule.
As Neptune only targets Triton as its optimizer, it cannot make use of
features on newer GPUs such as asynchronous tensor core matmul,
and delivers subpar performance compared to newer kernel libraries.

\textbf{Tensor Operator Fusion.}
In GPU kernels, multiple outermost loops execute in parallel across different GPU thread blocks.
Combining compute operations under the same parallel loops can bring significant
performance benefit,
since it reduces data communication across thread blocks. This technique is called operator fusion.
Operator fusion is not always possible for computation with complex data dependency patterns.
Neptune generalizes operator fusion to support fusing multiple reduction operations,
and provides them as program transformation passes that \NAME{} uses.

}

\section{\NAME{} Auto-Scheduler}
\label{sec:auto-sched}
{The \NAME{} auto-scheduler finds high-performance schedules for an input program without
user intervention.
From the input program, it sketches a set of promising schedules, which we call
\textit{schedule seeds},
and refines decision parameters in these schedules using auto-tuning.
Seeds are generated by applying a sequence of \emph{scheduling passes}
on parts of the program in a specific order,
and each pass applies one class of optimizations, such as tiling or fusion.

The auto-scheduler is novel compared to previous auto-schedulers in three ways: 
(1) it is entirely built around targeting tile-based languages and takes important
steps to ensure compatibility with them, (2) it makes use of new advanced reduction fusion
techniques to attain deeper fusion than previous loop-based language auto-schedulers, and (3)
it employs heuristics to aggressively cut down on the search space, in order to allow for
auto-tuning in the minutes range.


\subsection{Auto-scheduling Algorithm}
The auto-scheduler builds up schedule seeds by applying scheduling passes
and enumerating compute statements in the program.
The input to the auto-scheduler is a program in which all compute statements
are perfectly nested in their own loops.
Each pass takes a schedule to build on and a compute statement it is currently transforming,
and produces one or multiple schedules,
if there are multiple promising ways to transform the program.
The auto-scheduler visits compute statements (nodes in the block graph) in a pre-order traversal,
producers to consumers, while maintaining a list of current schedules to apply passes on.
This process repeats until the auto-scheduler reaches the end of the block graph,
at which point all passes have been applied to all compute statements.
The list of current schedules then becomes the schedule seeds,
which contains design parameters (e.g., block tiling sizes) for the auto-tuner to further explore.

\subsection{Auto-Scheduler Transformations}
Our auto-scheduler applies four families of transformations:
tiling, fusion, cache management, and dataflow management.
These transformations capture the main steps needed to translate
and optimize a tensor-expression operator into a tile program ready for \NAME{}'s tile optimizers.

\subsubsection{Bi-level Tiling.}
The bi-level tiling transformation applies to a perfect loop nest,
splitting each loop into an inner loop and an outer loop,
and reordering them so that all outer loops are outside of all inner loops.
For example, a loop nest with two loops \texttt{i, j} turns into the loops
\texttt{i0, j0, i1, j1}, with the first two loops being the outer loops.
The other passes in the auto-scheduler transform only the outer loops;
the inner loops remain perfectly nested so they cleanly map to tiles in the tile IRs later.
Later, outer loops that are parallelizable (no data dependency across iterations)
are mapped to GPU blocks.
A size limit on inner loops is also enforced to prevent tiles from becoming
too large and not fitting into the cache - in this case, the respective large inner loop is itself
tiled into an inner and outer loop if possible.

\subsubsection{Operator Fusion.}
The auto-scheduler applies fusion transformations to fuse loop nests together,
to reduce the number of kernels produced and improve data locality.
There are two fusion strategies in the auto-scheduler:
classic loop fusion based on data dependency,
and a ``rolling update'' technique that enables fusion between reductions.
Rolling update is an advanced fusion technique from Neptune \cite{neptune}
which is beyond the capabilities of classic loop fusion,
fusing multiple reductions into an ``online'' kernel.
While classic loop fusion preserves data dependency,
rolling update intentionally breaks some dependency constraints
and repairs the program by changing compute expressions.

In the auto-scheduler, classic loop fusion and rolling update presents a trade-off:
rolling update enables more fusion opportunities but introduces a ``repair factor'' into the program,
which is extra computation with potential overhead.
For each loop nest $L$ to be fused, the auto-scheduler selects a fusion strategy to apply
and a destination loop $l$ to fuse into.
The destination $l$ is selected from all loop nests that precede $L$ in the block graph.
We employ a greedy heuristic for this decision, preferring to fuse into deeper loops in a loop nest.

A key challenge in auto-scheduling rolling update fusion is that rolling update is not a ``local'' transformation:
rolling update always applies to two reductions,
fusing them together along with all element-wise operations between them on the block graph.
Thus, our auto-scheduler
needs to employ a look-ahead strategy to decide whether such a nest will be able to fuse with a reduction nest. We discuss
the intricacies of this algorithm in \ref{sec:adv-fusion}.



\subsubsection{Data Localization.}
Data localization optimizations place data tensors on an inner (more local) memory level
(e.g.~shared memory and registers compared to global)
with the goal of reducing slow global memory traffic.
Initially, all tensors are located in global memory before this set of optimizations applies.
The auto-scheduler performs two types of data localization:
(1) it creates a copy of a tensor to a more local memory level
if the loop nest exhibits temporal reuse of this buffer
(i.e., it is expected this copy will be accessed many times),
and (2) it replaces the tensor in an inner memory level
if all tensor accesses are local to a thread block.

The auto-scheduler combines both data localization strategies with an analysis
that considers the capacity of a memory level,
as inner memory levels (such as shared memory) are often smaller.
Tensors which have a long lifetime and are large in size may cause additional memory pressure on
local memory.
Our auto-scheduler performs tensor lifetime and size analysis to identify
large buffers with long live ranges that would otherwise occupy scarce local memory.
For buffers that are read multiple times and are inexpensive to recompute,
we apply rematerialization and live-range splitting,
recomputing the buffer near each use to reduce memory pressure and improve occupancy.

\subsubsection{Regularization for Tile Optimization.}
Our auto-scheduler applies several transformations to align the data dependency patterns
of the program to what tile languages expect,
as tile languages often expect a small set of tile computation patterns
and can fail to optimize others.
The auto-scheduler applies a split-scan-buffer transformation
that attempts to remove tensor access with ``scan'' dependency
(e.g., \texttt{a[i, j] = f(a[i - 1, j])}).
Scan dependencies can exist in the input program itself,
or arise from previous transformations, such as rolling update,
and existing tile optimizers often fail to optimize scan dependency
or produce significantly suboptimal kernels.

The auto-scheduler explores the choice of using different tile optimizers
by creating a schedule seed for each tile optimizer,
and setting compile parameters, such as the number of warps and the number of pipeline stages,
for the given tile optimizer.
This allows the auto-tuner to explore
different backends and decide which one is most performant (\S\ref{sec:opt-pipeline}).

\subsection{Auto-scheduling Advanced Fusion}\label{sec:adv-fusion}

The \NAME{} auto-scheduler's fusion pass consists of two fusion primitives - classical
fusion and rolling update. Each of these primitives operates on a compute statement
and a loop nest, fusing the compute statement into that loop nest. Each
has its own use cases, and they are used together to fuse the program as much as possible.

Classical fusion is the simpler of the two: it always operates on an element-wise loop
nest directly followed by the target compute statement to be fused, which can be either
element-wise or a reduction. When there are multiple depths of loop nest that classical
fusion can fuse with, the deepest level is selected by the auto-scheduler to improve locality.

When the loop nest being fused into is a reduction loop nest, the rolling update primitive
is required. Rolling update works differently to classical fusion in two ways: (1) it only
operates on reduction compute statements, and (2) the reduction compute statement does not
need to directly follow the loop nest being fused into --- they can be separated by element-wise
loop nests along the dependence path, which will all be ``captured'' and fused into the loop
as well. These differences mean that an element-wise statement $e$ cannot be directly fused into a
reduction loop nest $L$ --- the only way to do so is via finding a reduction statement $r$ that is a successor
to $e$ and fusing $r$ into $L$ with rolling update, capturing $e$.

Our auto-scheduler needs to take account of these rules when deciding how to perform fusion.
The fusion pass operates in a pre-order fashion. First, every compute statement $s$ visited is greedily
attempted to be fused into every loop nest that is a direct producer of it via classical fusion.
Then, if $s$ is a reduction, it also tries to fuse $s$ into all predecessor
reduction loop nests via rolling update. 

Because rolling update is unable to directly fuse element-wise statements into reduction loops,
the auto-scheduler may come across such a statement $s$ where it is unable to perform fusion on that
statement. There are two scenarios in this case --- (1) there is some successor reduction compute statement
that can be fused with a reduction loop predecessor of $s$, or (2) operator fusion cannot happen.
The auto-scheduler will look ahead in this case to try finding such a successor, and if it fails to,
it will choose to start a new kernel with $s$. Since rolling update is required to fuse any two reduction
loops together, when rolling update fusion is feasible it will always fuse deeper than classical fusion can.
Consider a scenario where there is a loop nest $L$ with an element-wise loop $e$ around a reduction loop $r$.
$L$ has an elementwise consumer compute statement of $c_1$, and $c_1$ has a reduction consumer statement $c_2$.
In this case, when analyzing $c_1$, it would only be able to directly fuse with $e$, via classical fusion.
However, it would also be able to fuse with the deeper loop $r$ via capture from rolling update fusion
between $c_2$ and $r$. In this scenario, the auto-scheduler would visit $c_1$ before $c_2$ as part of pre-order traversal.
Greedily fusing $c_1$ with $e$ during visiting of $c_1$ will not impede $c_1$ fusing with $r$ via capture during the
visiting of $c_2$, illustrating that a greedy approach to classical fusion does not impede rolling update
fusion in the auto-scheduler.

There are several additional considerations needed when deciding to fuse a compute
statement with a loop. First, occasionally a compute statement will have multiple
producers. In this case, it can be beneficial to fuse with multiple of them in the same
schedule. 
Second, when deciding to fuse with a loop, we must ensure that the loop is tiled, and if
it isn't, to tile it before fusion. When a compute statement is fused into a (tiled)
loop, its loop nest will automatically become tiled as a result of the fusion too.

\subsection{Auto-Tuner}\label{sec:auto-tuner}
\NAME{} uses MetaSchedule's auto-tuner in order to generate and profile more candidate
schedules in hopes of finding high-performance variants. The auto-tuner runs after the
auto-scheduler, taking as input the produced schedule seeds from the auto-scheduler.
For each input seed, the auto-tuner analyzes the schedule and extracts tunable
parameters, such as tile dimensions and compiler options, specific to each of the
different backends which \NAME{} uses (Section~\ref{sec:opt-pipeline}).
It then heuristically selects values for these parameters and generates multiple
candidate schedules. \NAME{} applies these schedules to the input program, runs
the resulting program and selects the ones with the lowest latency.


\setminted{breaklines=true, breakanywhere=true, fontsize=\footnotesize, fontfamily=lmtt,
xleftmargin=0pt, xrightmargin=0pt, framesep=2pt}
\begin{figure*}[ht]
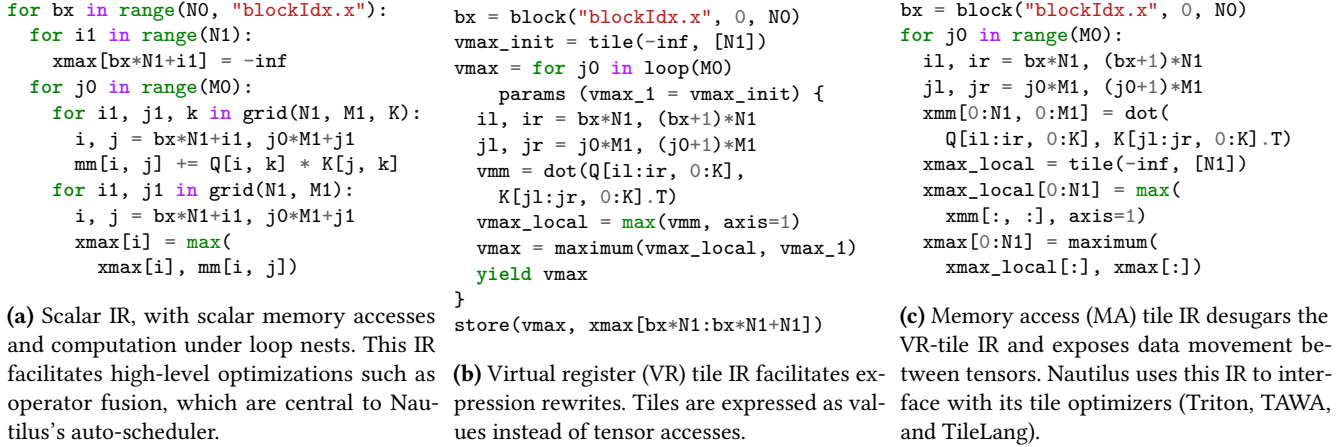

  \begin{subfigure}[b]{0.32\textwidth}
    \centering
    \begin{minted}{python}
for bx in range(N0, "blockIdx.x"):
  for i1 in range(N1):
    xmax[bx*N1+i1] = -inf
  for j0 in range(M0):
    for i1, j1, k in grid(N1, M1, K):
      i, j = bx*N1+i1, j0*M1+j1
      mm[i, j] += Q[i, k] * K[j, k]
    for i1, j1 in grid(N1, M1):
      i, j = bx*N1+i1, j0*M1+j1
      xmax[i] = max(
        xmax[i], mm[i, j])
    \end{minted}
    \caption{Scalar IR, with scalar memory accesses and computation under loop nests.
      This IR facilitates high-level optimizations such as operator fusion,
      which are central to \NAME{}'s auto-scheduler.}
  \end{subfigure}
  \hspace{2pt}
  \begin{subfigure}[b]{0.32\textwidth}
    \centering
    \begin{minted}{python}
bx = block("blockIdx.x", 0, N0)
vmax_init = tile(-inf, [N1])
vmax = for j0 in loop(M0)
    params (vmax_1 = vmax_init) {
  il, ir = bx*N1, (bx+1)*N1
  jl, jr = j0*M1, (j0+1)*M1
  vmm = dot(Q[il:ir, 0:K],
    K[jl:jr, 0:K].T)
  vmax_local = max(vmm, axis=1)
  vmax = maximum(vmax_local, vmax_1)
  yield vmax
}
store(vmax, xmax[bx*N1:bx*N1+N1])
    \end{minted}
    \caption{Virtual register (VR) tile IR facilitates expression rewrites.
    Tiles are expressed as values instead of tensor accesses.}
    \label{fig:vr-ir-example}
  \end{subfigure}
  \hspace{2pt}
  \begin{subfigure}[b]{0.32\textwidth}
    \centering
    \begin{minted}{python}
bx = block("blockIdx.x", 0, N0)
for j0 in range(M0):
  il, ir = bx*N1, (bx+1)*N1
  jl, jr = j0*M1, (j0+1)*M1
  xmm[0:N1, 0:M1] = dot(
    Q[il:ir, 0:K], K[jl:jr, 0:K].T)
  xmax_local = tile(-inf, [N1])
  xmax_local[0:N1] = max(
    xmm[:, :], axis=1)
  xmax[0:N1] = maximum(
    xmax_local[:], xmax[:])
    \end{minted}
    \caption{Memory access (MA) tile IR desugars the VR-tile IR and exposes data
      movement between tensors. \NAME{} uses this IR to interface with its tile
      optimizers (Triton, TAWA, and TileLang).
    }
  \end{subfigure}
  \caption{An example program in \NAME{}'s three IRs.}
  \label{fig:3ir-example}
  \vspace{-.1in}
\end{figure*}

\section{\NAME{}'s Successive Lowering Pipeline}
\label{sec:opt-pipeline}


Our optimization pipeline takes an input program and successively lowers it through a
scalar IR and two tile IRs.
While these types of IRs are commonly used in tensor compilers,
one of the two tile IRs in \NAME{} is unique: the virtual register tile IR (VR-tile IR),
necessitated by the need for \emph{expression rewriting at tile level}.
The other tile IR in \NAME{}, the memory access tile IR (MA-tile IR),
enables \emph{tile optimization}.
It is purposefully similar to the tile IRs that many tile compilers use,
simplifying interfacing with existing tile optimizers.

\textbf{Tile IRs.}
Figure \ref{fig:3ir-example} shows an example program in both tile IRs,
and scalar IR for reference.
In both tile IRs, tensor computation is built up from expressions over tiles.
The MA-tile IR exposes tile data movement as explicit tensor accesses,
which is typical in tile IRs of existing tile compilers.
\NAME{}'s VR-tile IR differs in that it expresses tiles as local variables (``virtual registers'')
rather than as memory accesses.
This design reduces data movement instructions that would otherwise clobber expression manipulation.
For example, \texttt{vmm}, \texttt{vmax\_local}, and \texttt{vmax} in Figure \ref{fig:vr-ir-example}
are each bound to a tile expression, so expression rewrites view them as one
expression when needed and freely move terms across them.

While eliminating all data movement instructions is a non-goal,
it can be beneficial to express more parts of the program as computations over virtual
tile registers.
We introduce a new loop construct to the VR-tile IR: the for-loop expression,
which has loop-carried parameters and returns values.
In Figure \ref{fig:vr-ir-example}, the loop expression \texttt{for j0 in loop(M0)}
takes parameter \texttt{vmax\_1} and returns \texttt{vmax}.
This design is in spirit similar to the memory-to-register (``mem2reg'')
found in some general-purpose CPU-centric compilers such as LLVM.
Structural for-loops express value evolution over loop iterations natively
instead of as individual update-writes to a tensor
and enable powerful rewrites over loop iterations.

\textbf{Expression Rewriting.}
\NAME{}'s expression rewriter applies a set of rewrite rules recursively on the VR-tile IR.
Table \ref{tab:rewrite-rules} lists rewrite rules that are specific to \NAME{}.
In addition to standard algebraic identities, we include rules over loop expressions,
and rules that are informed by GPU hardware (e.g. rewriting $e^x$ using $2^x$).
Thanks to the design of the VR-tile IR, simple rewrite rules are capable of
powerful rewrites over loop iteration, and it is easy to extend this pass with more rewrite rules.
For example, the second rule in Table \ref{tab:rewrite-rules}
cancels out division and multiplication over loop iterations, removing $2n - 1$
operations from the program.


\begin{table}[ht]
  \scriptsize
  \centering
  \caption{Representative rules in \NAME{}'s expression rewriter. Standard algebraic
  identities are used but not listed.}
  \tt\vspace{-.1in}
  \label{tab:rewrite-rules}
  \begin{tabular}{ll}
    \hline
    \textbf{Pattern} & \textbf{Replacement} \\
    \hline
    $ x = \text{For}(...) \{ \text{yield } c \cdot x \} $ &
    $ x' = \text{For}(...) \{ \text{yield } x \};\ x = c \cdot x' $ \\
    $ x = \text{For}(i: n) \{ \text{yield } $$x / y_i \cdot y_{i+1} \} $ &
    $ x' = y_n \cdot \text{For}(i: n) \{ \text{yield } x \} $ \\
    $ \exp(x) $ & $ \text{exp2}(\text{log2}(e) \cdot x) $ \\ \hline
  \end{tabular}
  \vspace{-.1in}
\end{table}

\textbf{Tile Optimizations.} \NAME{}'s MA-tile IR is the language where tile
optimizations happen.
It builds on the tile IR of Neptune, which Neptune uses to apply tile optimizations from Triton.
In \NAME{}, we design the MA-tile IR as a super set of multiple existing tile languages,
which enables \NAME{} to use multiple tile optimizers: Triton, Tawa~\cite{tawa}, and
TileLang~\cite{tilelang}.
The auto-scheduler includes the choice of tile optimizer as part of the schedule
(retained on the program during lowering),
and selects the best-performing tile optimizer for the input program and target hardware.
MA-tile contains plain for-loops and exposes tile load/store information,
which tile optimizers require to perform SIMD optimizations.
After expression rewriting, \NAME{} translates the program from the VR-tile IR into the MA-tile IR,
by desugaring the loop expression and tile-valued virtual registers in the VR-tile IR.
}

\vspace{-.08in}
\section{Implementation}
\vspace{-.04in}
\label{sec:impl}
{
\NAME{} reuses components from TVM~\cite{tvm}, Neptune~\cite{neptune}, and multiple tile-based compilers~\cite{triton,tawa,tilelang}.
We use TVM's tensor expression (TE) language as the input format of \NAME{}
to describe the input tensor program.
The scheduling language uses infrastructure from TVM's scheduling language for TensorIR.
The scalar IR, where scheduling happens, and the memory access tile IR are adapted from Neptune.
Including VR-tile IR which is a \NAME{} creation, the implementations of all three IRs
use TVM TensorIR {code infrastructure as a foundation.}

For the \NAME{} auto-scheduler, 
we developed a custom design-space generator 
to create our pre-order pass traversal. We implement each of our passes as a TVM MetaSchedule rule which is fed to this design space generator.
%
The auto-scheduler uses an
evolutionary search algorithm with a learned cost model.
For each schedule seed, it generates different schedules by varying tunable parameters.
It then predicts the performance of each schedule with the cost model, and
runs the projected-best schedules on the target GPU.
The auto-tuning budget of \NAME{}'s auto-scheduler is measured by the number of empirical GPU
measurements.

\NAME{} has three tile-based compilers as back-ends and tile optimizers: Triton, Tawa~\cite{tawa}, and TileLang~\cite{tilelang}.
The auto-scheduler includes the choice of tile optimizer as part of the schedule,
and selects the best-performing tile optimizer for the input program and target hardware.
\NAME{} provides translators from its MA-tile IR to the tile-based language of each compiler,
and a runtime that manages kernel compilation, caching, and invocation through a unified interface.
}

\section{Experimental Methodology}
\label{sec:method}
{\textbf{Hardware Platforms.}
We evaluate \NAME{} on two hardware platforms:
(1) NVIDIA GH200 Grace Hopper Superchip, with an NVIDIA H100 GPU (Hopper arch.);
(2) a workstation with an NVIDIA RTX 5090 GPU (Blackwell arch.).

\begin{table}[b]
  \small\vspace{-.1in}
  \centering
  \caption{LLM operators used in our evaluation. }
  \label{tab:operators}
  \begin{tabular}{l|l|l} \Xhline{2\arrayrulewidth}
    \textbf{Operator} & \textbf{Description}    & \textbf{Base Arch.} \\ \hline
    Global      & Global Attention        & ViT 1.2B \\
    Causal      & Causal Attention (MHA)  & \makecell[l]{Llama 2 7B, Qwen2 7B, \\ Qwen3 8B} \\
    GQA         & Grouped Query Attention & GLM-4 9B \\
    \Xhline{2\arrayrulewidth}
  \end{tabular}
\end{table}

\textbf{End-to-End Inference Experiments.}
We evaluate \NAME{} on the large language models (LLM)
and vision language models (VLM) shown in Table~\ref{tab:operators}.
We distinguish two steps of LLM inference in our evaluation: prefill (PF) and decoding (DC).
Prefill consumes the entire prompt in parallel to build a key-value cache,
while decoding generates one token at a time using the cache.
We run each model in both prefill and decoding modes on varying input batch sizes (1 and 8),
sequence lengths ($2^{10} = 1024, \dots, 2^{15} = 32768$),
and data precisions (FP16 and FP8-E4M3).

\textbf{Operator Experiments.}
Table~\ref{tab:operators} presents three individual operators that \NAME{} optimizes.
We extracted them from the transformer-based models used in our end-to-end inference experiments.
The input batch size, sequence length, and data precision are the same as in the end-to-end
experiments.
We refer to the tuple of hardware platform, operator, input shape, and input precision as
a \textbf{setup}.
For each setup, we run \NAME{}'s auto-tuner for 256 empirical measurements.

\textbf{Baselines.}
We compare \NAME{} to a comprehensive set of 12 baselines listed here, with underscores
marking the names we refer to them.
Our baselines include six tensor compilers:
\uline{Tawa} 3.3~\cite{tawa}, \uline{TileLang} 0.1.8~\cite{tilelang}, Triton 3.6.0,
\uline{Helion} 0.3.1,
FlexAttention (\uline{FlexAttn}) in \uline{PyTorch} 2.11.0, and \uline{TVM} 0.18.
Triton, Tawa, and TileLang are tile-based compilers.
We use the stock attention implementation for Tawa and TileLang,
and two mainstream implementations for Triton:
\uline{OpenAI Triton} \cite{openai-fused-attn} and \uline{Tri-Dao Triton} \cite{flash-attn-repo}.
We compare five optimized kernel libraries and header-only libraries:
\uline{ThunderKittens}~\cite{thunderkittens} (obtained from TK authors),
\uline{PyTorch} 2.11.0 \cite{pytorch}, \uline{cuDNN} 9.11.0,
\uline{Tri-Dao} 2.8.3 (Dao-AILab's CUTLASS implementation) \cite{flash-attn-repo},
and \uline{FlashInfer} 0.6.7 \cite{flashinfer}.
Each baseline supports a subset of the setups we evaluate.
For PyTorch and FlexAttn, we use \texttt{compile(mode=`max-autotune', backend=`inductor')} for high performance.
For Tawa and TileLang, we use implementations following the FlashAttention-3 strategy. We do not evaluate FlashAttention-4 \cite{zadouri2026flashattention4algorithmkernelpipelining} because it includes approximation techniques that are outside the scope of our exact-attention evaluation.

\textbf{Profiling and Performance Reporting.}
We use the CUDA Events API to measure the end-to-end inference latency of each model,
and we report the throughput in TFLOPs per second.
For the throughput, following~\cite{Megatron-LM}, we calculate the theoretical
floating-point operations (FLOPs) required for the attention computation
based on the model configuration (such as sequence length, number of heads, and hidden
dimension) and divide it by the measured latency.
In operator experiments, we measure kernel latencies with Nsight Systems
(\texttt{nsys}) on Nvidia GPUs, for better precision on short-duration kernels.
We report \textbf{\textit{speedup}} of \NAME{} over the best baseline:
if \NAME{}'s implementation has mean latency $t_0$ and the baselines have $t_1, \dots t_n$,
then our speedup is $\min(t_1, \dots, t_n) / t_0$.
For all experiments, we run the workload once to warm up and discard the
result, and repeat the measurement 10 times to report the mean latency.
To average these metrics across setups, we use geometric means.
}

\section{Evaluation}
\label{sec:eval}
{


\noindent\textbf{RQ1:}
Can \NAME{} accelerate end-to-end transformer-based model inference and outperform existing
tensor compilers and attention-specialized inference frameworks?

\begin{figure*}[t!]
  \centering
  \includegraphics[width=0.9\textwidth]{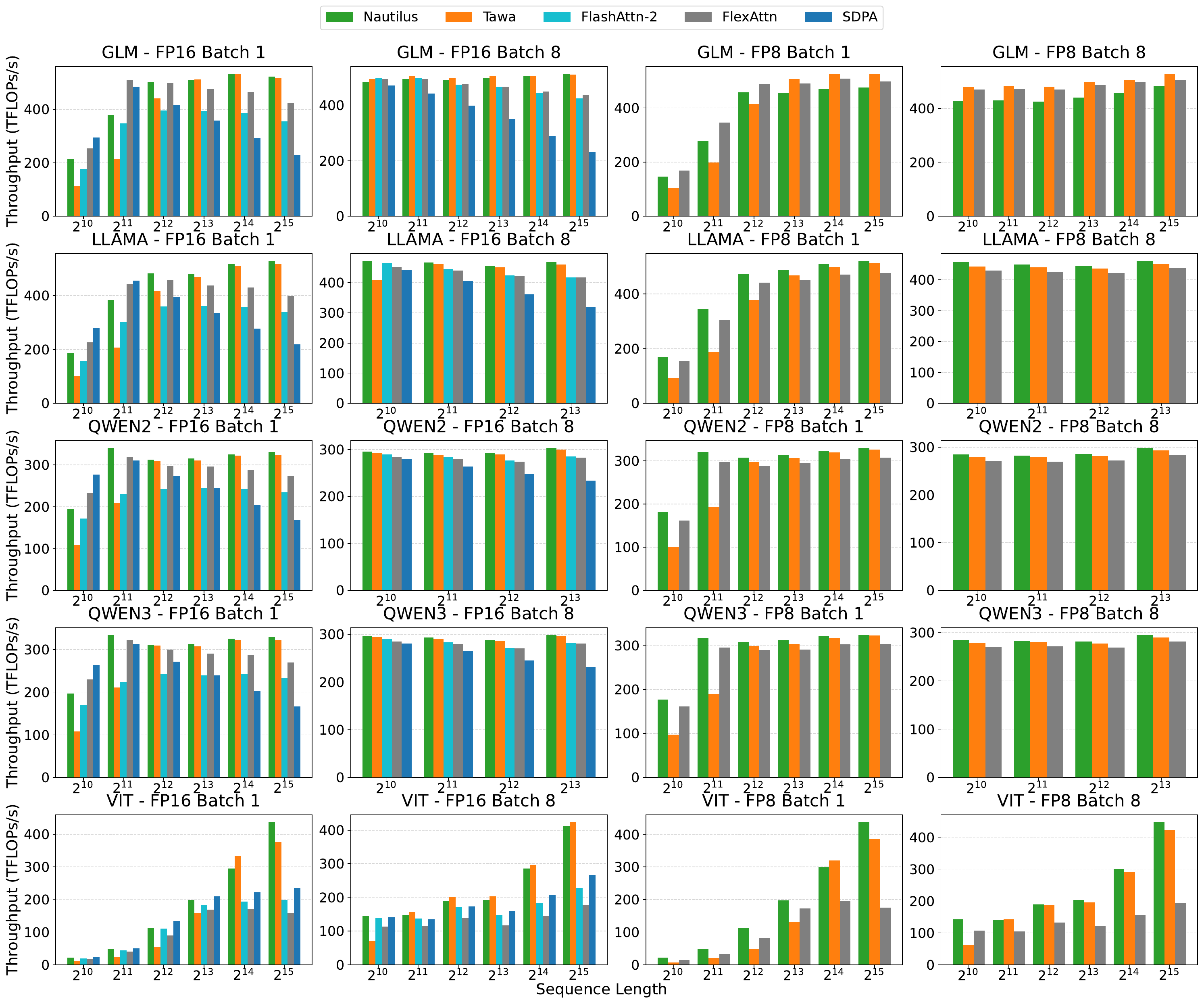}
  \vspace{-1em}
  \caption{Throughput (TFLOPs/s) on GH200 across different models.
    We evaluate \NAME{} against baselines across five foundation models (GLM,
    Llama, Qwen2, Qwen3, ViT).
    The benchmarks cover FP16 and FP8 precisions with Batch Size 1 and 8, for sequence
  lengths ranging from 1k to 32k. }
  \label{fig:gh200-throughput}
\end{figure*}

\begin{figure*}[t!]
  \centering
  \includegraphics[width=0.95\textwidth]{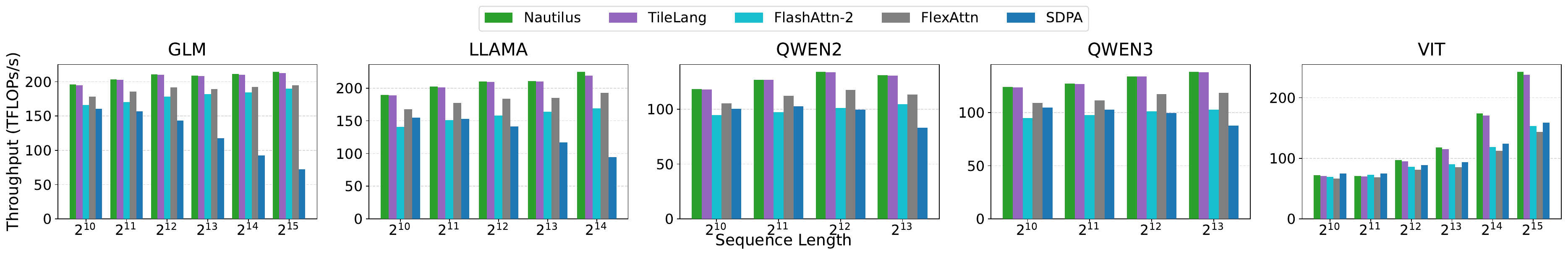}
  \vspace{-1em}
  \caption{Throughput (TFLOPs/s) on RTX 5090 across different models.
    We evaluate \NAME{} against baselines across five foundation models (GLM,
    Llama, Qwen2, Qwen3, ViT).
    The benchmarks cover sequence lengths ranging from 1k to 32k.
  }
  \label{fig:5090-throughput}
\end{figure*}

\noindent\textbf{RQ2:}
Can \NAME{} generate kernels that surpass the performance of kernels from existing tensor compilers?

\noindent\textbf{RQ3:}
Can \NAME{} generate kernels that have comparable performance to manually optimized
kernels in inference frameworks, which represent today's state of the art?

\noindent\textbf{RQ4:}
Ablation studies: How much do components of \NAME{} contribute to the performance of
\NAME{}'s attentions?

\subsection{End-to-End Model Inference}
\label{sec:eval-e2e}

\begin{table}[b!]
  \centering
  \small
  \setlength{\tabcolsep}{2.5pt}
  \caption{Geomean of relative throughput of \NAME{} against baselines on GH200
  across sequence lengths and batch sizes.}
  \label{tab:rq1-summary}
  \def\dgbox{\diagbox[width=6em, height=2.5em]{Model}{Method}}
  \renewcommand{\arraystretch}{0.93}
\vspace{-.07in}
  \begin{tabular}{l|cccccc}
    \toprule
    \dgbox & Flash-2 & Flex & SDPA & Tawa & Flex-FP8 & Tawa-FP8 \\
    \midrule
    GLM     & 1.17 & 1.03 & 1.33 & 1.11 & 1.07 & 1.16 \\
    LLAMA   & 1.23 & 1.06 & 1.26 & 1.18 & 1.08 & 1.17 \\
    QWEN2   & 1.20 & 1.05 & 1.20 & 1.12 & 1.07 & 1.13 \\
    QWEN3   & 1.21 & 1.05 & 1.21 & 1.12 & 1.06 & 1.13 \\
    VIT     & 1.28 & 1.54 & 1.15 & 1.28 & 1.59 & 1.43 \\
    \bottomrule
    \textbf{Average} & 1.22 & 1.13 & 1.23 & 1.16 & 1.16 & 1.20 \\
    \bottomrule
  \end{tabular}\vspace{-.03in}
\end{table}

\textbf{Overall Trends.}
To answer RQ1, we evaluate \NAME{} on two hardware generations: GH200 (Hopper) and RTX
5090 (Blackwell), using five representative foundation models:
GLM~\cite{du2022glmgenerallanguagemodel}, Llama2~\cite{touvron2023llama2openfoundation},
Qwen2~\cite{yang2024qwen2technicalreport}, Qwen3~\cite{yang2025qwen3technicalreport}, and
ViT~\cite{dosovitskiy2021imageworth16x16words}.
Table~\ref{tab:rq1-summary} and Table~\ref{tab:rq1-summary-5090} present the averaged
relative throughput of \NAME{} relative to state-of-the-art baselines including both
vendor-optimized libraries (FlashAttention-2, SDPA) and compiler-based approaches
(FlexAttn, Tawa~\cite{tawa}), where each cell represents the geometric mean over varying
sequence lengths and batch sizes.

Across the five representative foundation models in Table~\ref{tab:rq1-summary}, \NAME{}
achieves consistent performance gains in both FP16 and FP8 regimes.
In the FP16 setting, \NAME{} achieves an average speedup of $1.22\times$ over
FlashAttention-2 and $1.23\times$ over SDPA.
The performance advantage is most pronounced on the ViT architecture, where \NAME{}
achieves up to $1.54\times$ speedup against FlexAttn.
In the FP8 precision, \NAME{} shows improvements with an average speedup of $1.16\times$
over FlexAttn (FP8) and $1.20\times$ over Tawa (FP8).

On the RTX 5090 platform (Table~\ref{tab:rq1-summary-5090}), \NAME{} continues achieving
higher relative throughput.
In the FP16 setting, \NAME{} achieves an average relative throughput of $1.26\times$ over
FlashAttention-2 and $1.42\times$ over SDPA.
Compared to the TileLang baseline, \NAME{} shows a slight improvement, since \NAME{}
automatically selects TileLang as the optimal backend on the RTX 5090 architecture, and
uses the auto-scheduler and auto-tuner to improve kernel performance \mbox{beyond the
default configurations.}

\begin{table}[t!]
  \centering  \vspace{-.02in}

  \small
  \setlength{\tabcolsep}{6.5pt}
  \caption{Geometric mean of relative throughput of \NAME{} against baselines on RTX 5090
  across sequence lengths.}
  \label{tab:rq1-summary-5090}
  \def\dgbox{\diagbox[width=6em, height=2.5em]{Model}{Method}}
  \renewcommand{\arraystretch}{0.93}
  \vspace{-.08in}
\begin{tabular}{l|cccc}
  \toprule
  \dgbox & Flash-2 & Flex & SDPA & TileLang \\
  \midrule
  GLM     & 1.16 & 1.10 & 1.74 & 1.01 \\
  LLAMA   & 1.32 & 1.14 & 1.59 & 1.01 \\
  QWEN2   & 1.28 & 1.14 & 1.33 & 1.00 \\
  QWEN3   & 1.32 & 1.15 & 1.33 & 1.00 \\
  VIT     & 1.23 & 1.30 & 1.18 & 1.02 \\
  \midrule
  \textbf{Average} & 1.26 & 1.16 & 1.42 & 1.01 \\
  \bottomrule
\end{tabular}
  \vspace{-.2in}

\end{table}

Figure~\ref{fig:gh200-throughput} and Figure~\ref{fig:5090-throughput} report the
detailed throughput for these models across varying sequence lengths
($1k$ to $32k$) and batch sizes, where the y-axis shows the throughput in TFLOPs per second.
\NAME{} consistently outperforms baselines because its auto-scheduler and auto-tuner
identify optimized parameters for every specific workload, whereas baselines rely on
static rules/configs.

Similarly, \NAME{} achieves greater throughput gains on ViT than on LLMs.
This is because ViT involves less common attention configurations, with fewer
layers, fewer attention heads, and smaller hidden dimensions, which the
baselines often do not optimize well. \NAME{}, instead, generates efficient
kernel optimizations for these workloads.

\begin{figure*}[t!]
  \centering
  \begin{subfigure}[b]{0.95\textwidth}
    \includegraphics[width=\textwidth]{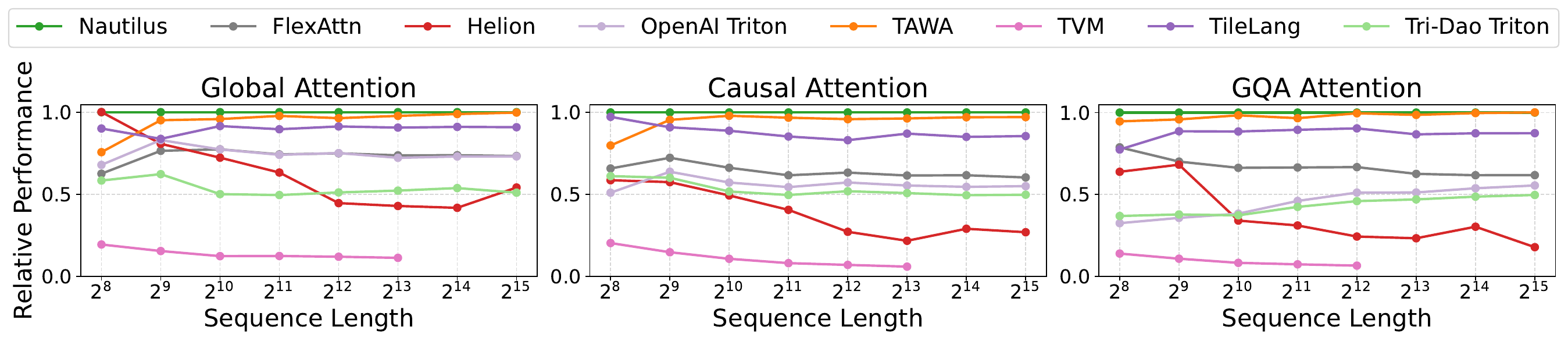}
    \vspace{-1.7em}
    \caption{GH200: FP16 performance (higher is better).}
    \label{fig:gh200-fp16-tc}
  \end{subfigure}

  \begin{subfigure}[b]{0.95\textwidth}
    \includegraphics[width=\textwidth]{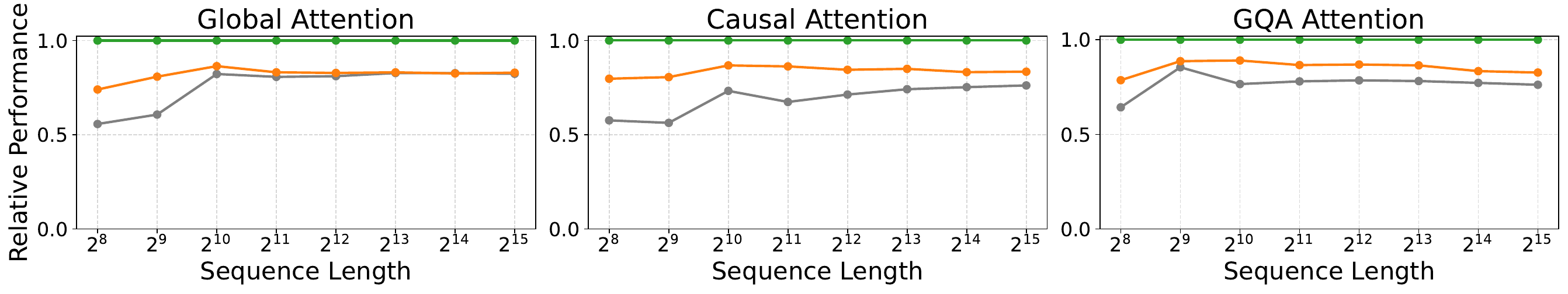}
    \vspace{-1.7em}
    \caption{GH200: FP8 performance (higher is better).}
    \label{fig:gh200-fp8-tc}
  \end{subfigure}

  \begin{subfigure}[b]{0.95\textwidth}
    \includegraphics[width=\textwidth]{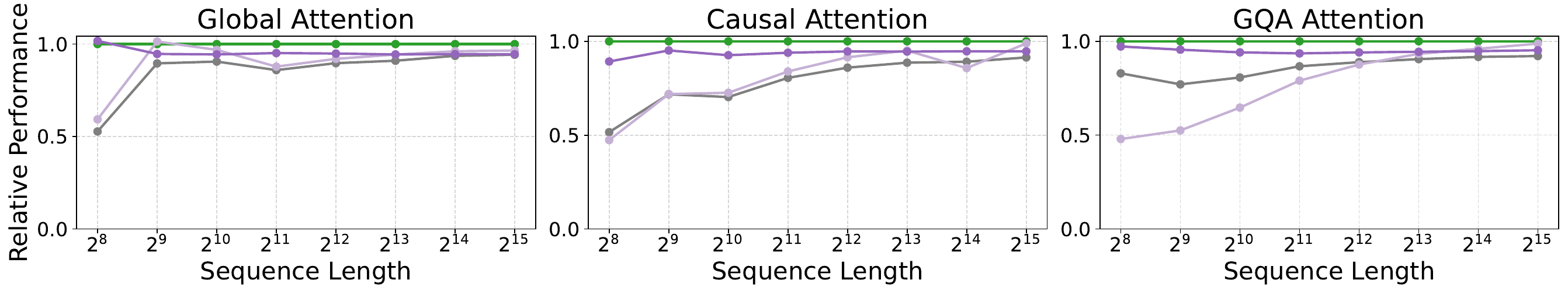}
    \vspace{-1.7em}
    \caption{RTX 5090: FP16 performance (higher is better).}
    \label{fig:rtx5090-fp16-tc}
  \end{subfigure}

  \caption{Performance of \NAME{} kernels vs. kernels by other tensor compilers on GH200
    and RTX 5090 platforms.
  The y-axis shows relative performance of all compilers normalized to \NAME{} for each setup.}
  \vspace{-.1in}
  \label{fig:compiler-prefill-ablation}
\end{figure*}

\subsection{Detailed Kernel-level Latency Evaluation}

Figure~\ref{fig:compiler-prefill-ablation} illustrates the relative performance of
\NAME{} against leading tensor compilers across three attention-based operators on GH200
and RTX 5090 platforms.
Each subfigure reports the performance of baselines normalized to \NAME{} (y-axis) over
varying input sequence lengths (x-axis).
On the GH200 platform, \NAME{} achieves the best performance in the majority of setups.
At short sequence lengths, our auto-scheduler and auto-tuner gain a significant advantage
over Tawa by automatically selecting the faster TileLang~\cite{tilelang} backend and
tuning parameters towards smaller block sizes and fewer pipeline stages.
For instance, on short sequences, \NAME{} selects a lighter $64 \times 128$ tile size
with 4 warps, while Tawa defaults to a heavy $128 \times 128$ size with 8 warps.
This adaptability provides even larger benefits for FP8 precision. \NAME{} adjusts
parameters such as pipeline stages from 1 to 4 based on sequence length which the
performance of models in FP8 relies heavily on, so \NAME{} can achieve bigger performance
improvements compared to FP16.
On the RTX 5090, \NAME{} consistently outperforms baselines, with larger speedups
observed on longer sequences where our autoscheduler selects the rolling update
optimization to overcome TileLang.
In FP8 precision setups (Figure \ref{fig:gh200-fp8-tc}), \NAME{} demonstrates a
significant advantage, achieving approximately 20\% improvement on average over the
strongest baselines, enabled by the optimized configurations identified by our auto-tuner.
Our auto-tuner shows hardware adaptability by adjusting the search space parameters, e.g., block sizes, to accommodate the limited shared \mbox{memory of the RTX 5090.}

We do not show Neptune results since Neptune's performance is limited by Triton's performance,
due to being tied to Triton for code generation, which struggles to generate efficient code for these
operators on Hopper and Blackwell.
Our measurements show Neptune's speed is 0.5x-0.8x of that of Tawa or TileLang. 
This result indicates that the main bottleneck is backend code generation rather than schedule quality. Our results suggest that the schedules found by \NAME{} are already competitive with the manually written schedules used by the Neptune developers, and that the remaining gap comes from Triton's lower code quality on these platforms.





\subsection{\NAME{} vs Manually-Written Kernels}


Appendix~\ref{sec:vendor-libs} compares \NAME{} against heavily optimized,
hand-written vendor libraries on GH200 and RTX 5090.
Across both hardware generations, \NAME{} matches or exceeds the performance of
state-of-the-art vendor libraries in most setups, and remains close to the highly
optimized cuDNN kernels in the remaining cases.



\subsection{Numerical Stability}
In Appendix~\ref{sec:num-stab}, we evaluate the numerical behavior of \NAME{}-generated kernels.
Using realistic Qwen2.5 inputs and an FP64 attention implementation as reference, \NAME{} shows numerical error comparable to Tri-Dao Attention and FlexAttention across PF Global, PF Causal, and PF GQA.
Averaged across the three operators, \NAME{} achieves an RMS absolute error of $4.96\times 10^{-5}$, compared with $4.90\times 10^{-5}$ for Tri-Dao and $5.02\times 10^{-5}$ for FlexAttention.

\subsection{Ablation Studies}

\begin{table}[t]
  \centering
  \caption{Ablation study of \NAME{}'s latency}
  \vspace{-.1in}
  \label{tab:ablation}
  \begin{tabular}{lc}
    \toprule
    \textbf{Configuration} & \textbf{Latency ($\mu s$)} \\
    \midrule
    \NAME{} (Full System) & 7.43 \\
    w/o Auto-tuning & 9.09 \\
    w/o Auto-tuning \& Expression Rewriting & 9.45 \\
    w/o Auto-scheduling & 10.93 \\
    \bottomrule
  \end{tabular}
  \vspace{-.1in}
\end{table}

To understand the importance of each component of \NAME{},
we perform ablation studies on the performance of \NAME{}'s kernels.
We select the Global attention operator with sequence length 256 on GH200.
Table \ref{tab:ablation} shows the results of this experiment.
The full \NAME{} system achieves the latency of 7.43 $\mu s$.
Disabling components progressively increases latency: removing the auto-tuner causes a
significant drop to 9.09$\mu$s. This shows the importance of parameter tuning.
Additionally disabling expression rewriting leads to further degradation to 9.45 $\mu s$.
Finally, the configuration with no auto-scheduling yields the highest latency of 10.93 $\mu s$.

\noindent\textbf{Autoscheduler Statistics.} \NAME{}'s auto-scheduling search is
relatively fast and finishes within a minute in most cases due to aggressive pruning. 
The auto-tuner takes a bit longer, taking from seconds to 10
minutes, as it needs to compile and runs many kernels.
\NAME-generated schedules contain a variety of different optimizations. GQA has several
reshape and broadcast operations added with respect to the base prefill attention, in
addition to an \mbox{extra loop nest.}
}

\section{\NA{Related Work}}
\label{sec:rw}
{
\textbf{LLM Inference Frameworks.}
A large body of work focuses on large language model (LLM) inference and serving,
including vLLM~\cite{vllm}, DeepSpeed Inference~\cite{deepspeed-inference},
NVIDIA FasterTransformer~\cite{fastertransformer}, and TensorRT-LLM~\cite{tensorrtllm}.
These systems primarily optimize at model and higher levels with techniques
like KV-cache management and request scheduling.
They typically do not provide their own kernels
and rely on kernel libraries to back the underlying tensor computation.
\NAME{} is a tensor compiler that automatically generates high-performance kernels and
integrates them into DNN inference pipelines.
Therefore, \NAME{} is complementary to these LLM-serving frameworks
and can serve as a backend that supplies high-performance kernels,
with performance and portability beyond what fixed \mbox{library kernels can provide.}

\textbf{Tile-based Programming Frameworks.}
Tile languages like Triton~\cite{triton}, TileLang~\cite{tilelang},
cuTile~\cite{cutile}, and others~\cite{cutlass,thunderkittens,tawa}
produces optimized mapping of tiles to warps and threads.
Tile languages are a low-level abstraction that leaves higher-level optimizations,
such as operator fusion and expression rewriting, on the table for programmers.
\NAME{} automates these high-level optimizations via auto-scheduling,
delegates tile optimizations to multiple tile compilers (Triton, Tawa, and TileLang),
and chooses between tile compilers dynamically based on performance.
\NAME{} is integrated with Triton, Tawa, and TileLang,
showing that its principle is general and can be applied to other tile compilers.

\textbf{Scheduling Tensor Compilers and Auto-Schedulers.}
Schedule-based tensor compilers such as Halide~\cite{halide}, TVM~\cite{tvm}, AKG~\cite{akg}, and
Tiramisu~\cite{tiramisu}, separate algorithm from optimization decisions (``schedule'').
Neptune~\cite{neptune} limits the schedule search space by delegating low-level
optimization to tile optimizers.
Auto-scheduling frameworks like Ansor~\cite{ansor}, MetaSchedule~\cite{metaschedule},
and FlexTensor~\cite{flextensor}
automate schedule generation and free users from manually supplying schedules.
\NAME{}'s novel scheduling language adapts to the optimization needs of modern GPUs,
and its novel auto-scheduler overcomes the challenges of complex
graph-level scheduling comprising advanced fusion strategies.

\textbf{Super-optimizing Tensor Compilers.}
Super-optimizing compilers such as TASO~\cite{taso} and Mirage~\cite{mirage}
transforms the given tensor program in ways that may not preserve semantic equivalence,
and impose additional checks to select correct implementations.
As an instance, Mirage is a hierarchical super-optimizer that
jointly transforms the program at thread, tile, and operator levels.
Mirage super-optimization produces a search space where search can take minutes to hours
even with pruning.
Mirage uses a probabilistic correctness test, producing kernels that are potentially unsound,
contrasting with \NAME{}'s guaranteed correctness.

\textbf{Pytorch-Based FlashAttention.}
Recently, Flashlight~\cite{flashlight} introduces a semantic fusion pass on top of the PyTorch 
TorchInductor compiler to generate FlashAttention-style fused kernels from PyTorch attention definition.
The main optimizations Flashlight supports are semantic fusion and limited tiling,
which are a set of graph rewrite rules specific to the TorchInductor IR.
Unlike Flashlight, which performs a limited set of optimizations and must generate Triton kernels
as it is bound to the PyTorch infrastructure,
\NAME{} provides a comprehensive optimization framework through its unique three-IR architecture.
\NAME{} combines operation-level optimizations, expression rewrites, and tile optimization in one pipeline,
and selects from multiple tile {optimizers based on performance.} \NA{Flashlight is not available as open-source and we could not experimentally compare the results.}

\textbf{Graph- and Model-level Optimizations.}
Graph compilers and model-level optimizers such as XLA~\cite{xla}, Glow~\cite{glow},
BladeDISC~\cite{bladedisc}, and TASO~\cite{taso} focus on optimizing entire computation graphs.
They perform operator fusion, layout transformation, and device placement, then lower
subgraphs to code generators or libraries.
These systems often skip over kernel-level optimizations,
relying on a separate kernel generator for the final code.
\NAME{} captures operator fusion as one of the most important graph-level optimizations,
and integrates it with kernel-level optimizations,
enabling their joint optimization in the space of schedules.

\textbf{Algebraic Expression Rewriting.}
Expression rewriting is another dimension of tensor program optimizations,
where optimizers exploit algebraic identities to transform compute expressions in the
program and improve performance.
Expression rewriting often happens at graph level, such as in TASO~\cite{taso},
TenSat~\cite{tensat}, and XLA's algebraic simplifier~\cite{xla},
where expressions describe computation over whole tensors.
Existing compilers and languages such as TVM, MLIR~\cite{mlir} and Lift~\cite{lift}
allow expression rewriting at lower levels.
\NAME{}'s IR design enables expression rewriting over tiles,
and \NAME{} rewrite rules that already demonstrate its effectiveness.
\NAME{}'s IR can be integrated with general-purpose rewriting engines, such as egg~\cite{egg},
to provide {powerful expression rewriting capabilities.}

}

\section{\NA{Conclusion}}
\label{sec:conclusion}
{We have presented \NAME{}, a fully-automated GPU kernel compiler with a novel auto-scheduler that supports advanced operator fusion. And
\NAME{} presents a sequence of lowering IRs that enables to automatically deploy
tile-based frameworks as backends.
\NAME{}-generated attention kernels show up to 23\% improvement over state-of-the-art
compiler systems on NVIDIA GH200 and up to 42\% on NVIDIA RTX 5090. As hardware grows more complex and exposes more features to software, we believe compilers which
do the heavy optimization work like \NAME{} grow increasingly critical for enabling broader progress on these platforms.
}

\begin{acks}
We are grateful to Minsoo Kim for assistance with prototyping \NAME{}'s Tilelang adaptor
and comments on an earlier draft of this paper.
\end{acks}

\bibliographystyle{ACM-Reference-Format}
\bibliography{references}

\clearpage
\onecolumn
\appendix
\label{app}
\clearpage

\section*{Appendix}
\vspace{.3in}

\section{\NAME{} vs Vendor Libraries}
\label{sec:vendor-libs}

Appendix~\ref{sec:vendor-libs} provides the full comparison of Nautilus against
heavily optimized vendor libraries on GH200 and RTX 5090,
including detailed results for different attention variants and
sequence lengths.

\begin{figure*}[hb!]
  \centering
  \begin{subfigure}[b]{0.95\textwidth}
    \includegraphics[width=\textwidth]{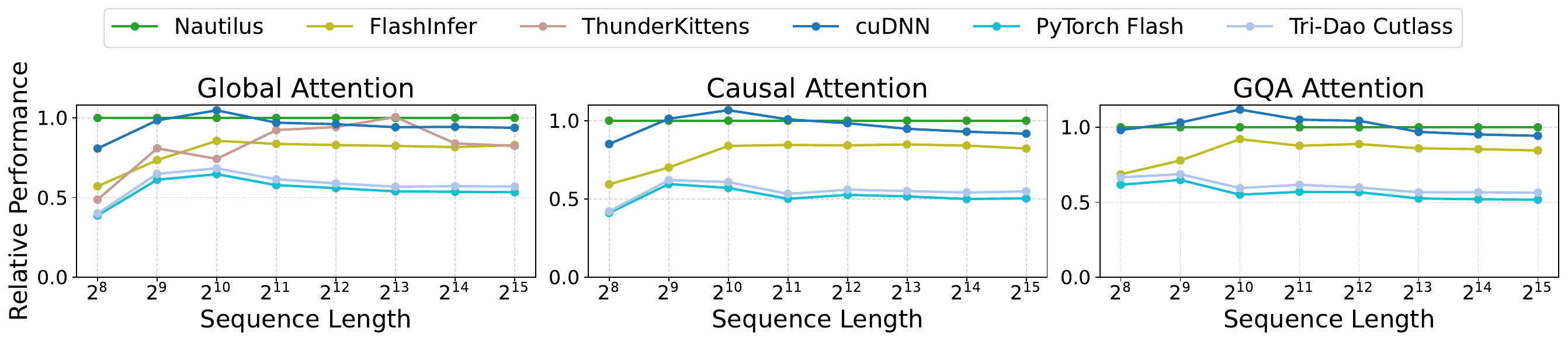}
    \vspace{-1.7em}
    \caption{GH200: FP16 performance (higher is better).}
    \label{fig:gh200-fp16-libs}
  \end{subfigure}

  \begin{subfigure}[b]{0.95\textwidth}
    \includegraphics[width=\textwidth]{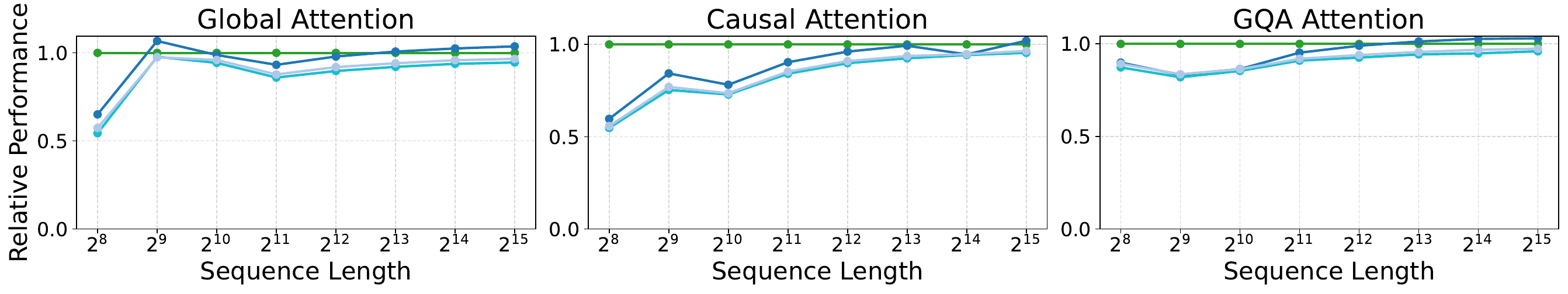}
    \vspace{-1.7em}
    \caption{RTX 5090: FP16 performance (higher is better).}
    \label{fig:rtx5090-fp16-libs}
  \end{subfigure}

  \caption{Performance of \NAME{} kernels vs. vendor libraries on GH200 and RTX 5090 platforms.
  The y-axis shows relative performance normalized to \NAME{} for each setup.}
  \label{fig:libs-prefill-comparison}
\end{figure*}

\clearpage

\section{Numerical Stability of \NAME{} Kernels}
\label{sec:num-stab}

\begin{wraptable}{r}{.5\textwidth}
\centering\small\vspace{.25in}
\caption{Numerical Error Summary (Absolute Error)}
\label{tab:numeric_error_absolute_error}
\begin{tabular}{llrrr}
\hline
Method & Variant & RMS & 90th \% & 99th \% \\
\hline
\multirow{3}{*}{Nautilus}
& PF Global & 4.9e-05 & 5.1e-05 & 5.5e-05 \\
& PF Causal & 5.0e-05 & 5.1e-05 & 5.3e-05 \\
& PF GQA    & 5.0e-05 & 5.1e-05 & 5.3e-05 \\
\hline
\multirow{3}{*}{Tri-Dao Attention}
& PF Global & 4.8e-05 & 5.0e-05 & 5.1e-05 \\
& PF Causal & 4.9e-05 & 5.0e-05 & 5.2e-05 \\
& PF GQA    & 4.9e-05 & 5.0e-05 & 5.2e-05 \\
\hline
\multirow{3}{*}{Flex Attention}
& PF Global & 5.0e-05 & 5.2e-05 & 5.4e-05 \\
& PF Causal & 5.0e-05 & 5.1e-05 & 5.3e-05 \\
& PF GQA    & 5.0e-05 & 5.1e-05 & 5.3e-05 \\
\hline
\end{tabular}
\vspace{.65in}
\end{wraptable}
Rewriting or re-associating floating-point operations can affect a kernel's numerical
behavior. To study this effect, we compare \NAME{}, Tri-Dao, and FlexAttention on
different attention operators. The results show that \NAME{} has competitive numerical
behavior. For all three systems, we use a common set of inputs and compare their outputs
to an FP64 unfused attention kernel baseline.
We keep the schedule and configuration of each kernel the same as in the performance
evaluation. To collect realistic attention inputs, we run a pretrained Qwen2.5 model
on 1000 WikiText sentences with sequence length 2048, and record the resulting input tensors.

\begin{figure}[h]
  \centering
\vspace{.2in}
  \begin{subfigure}[t]{0.32\textwidth}
    \centering
    \includegraphics[width=\linewidth]{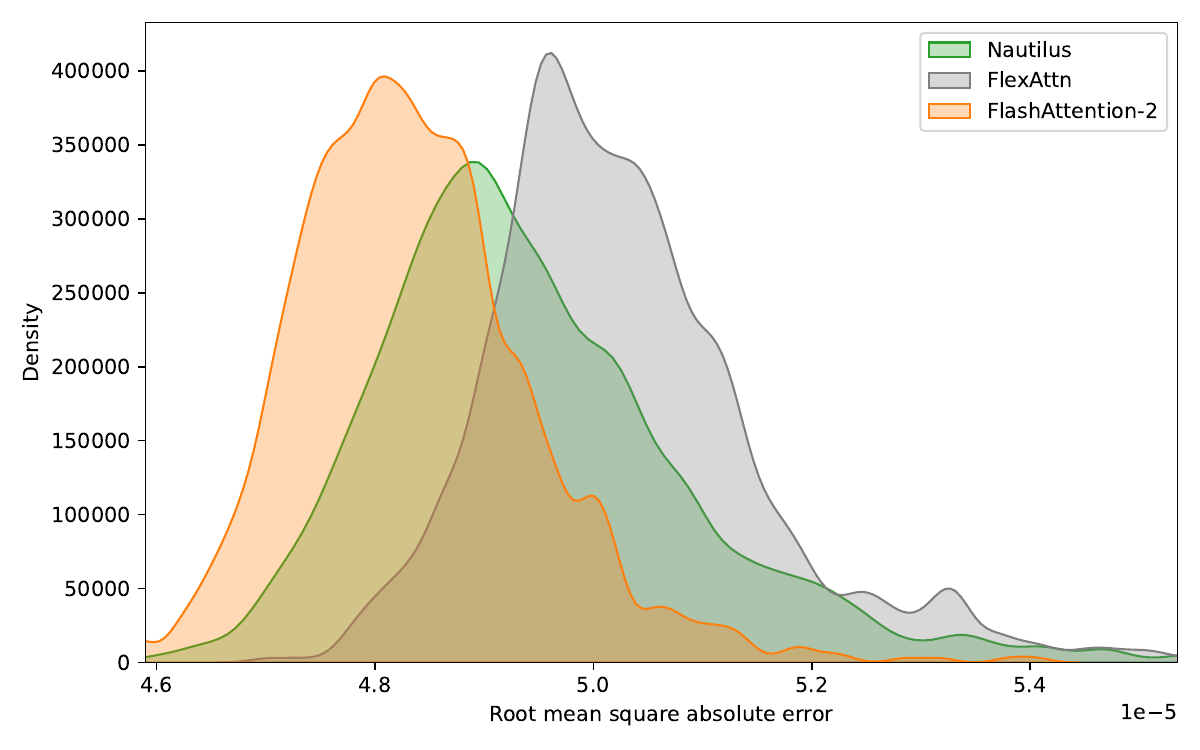}
    \caption{PF Global.}
    \label{fig:rmse-kde-global}
  \end{subfigure}
  \hfill
  \begin{subfigure}[t]{0.32\textwidth}
    \centering
    \includegraphics[width=\linewidth]{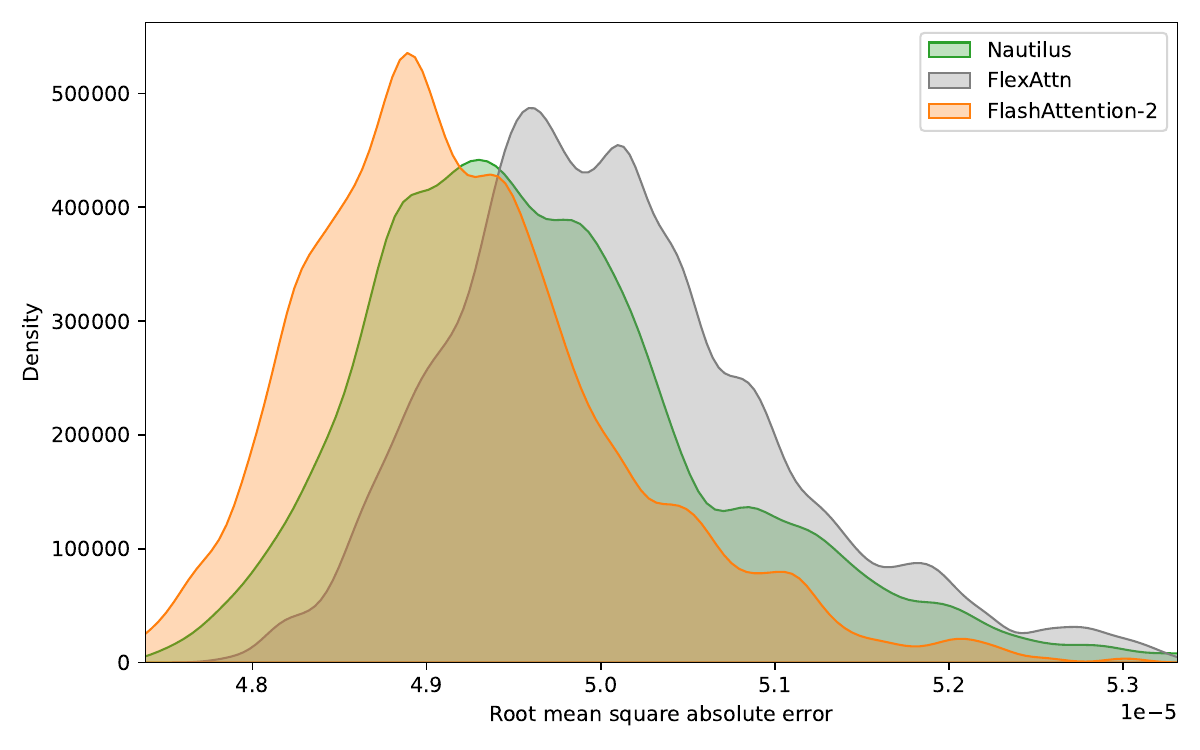}
    \caption{PF Causal.}
    \label{fig:rmse-kde-causal}
  \end{subfigure}
  \hfill
  \begin{subfigure}[t]{0.32\textwidth}
    \centering
    \includegraphics[width=\linewidth]{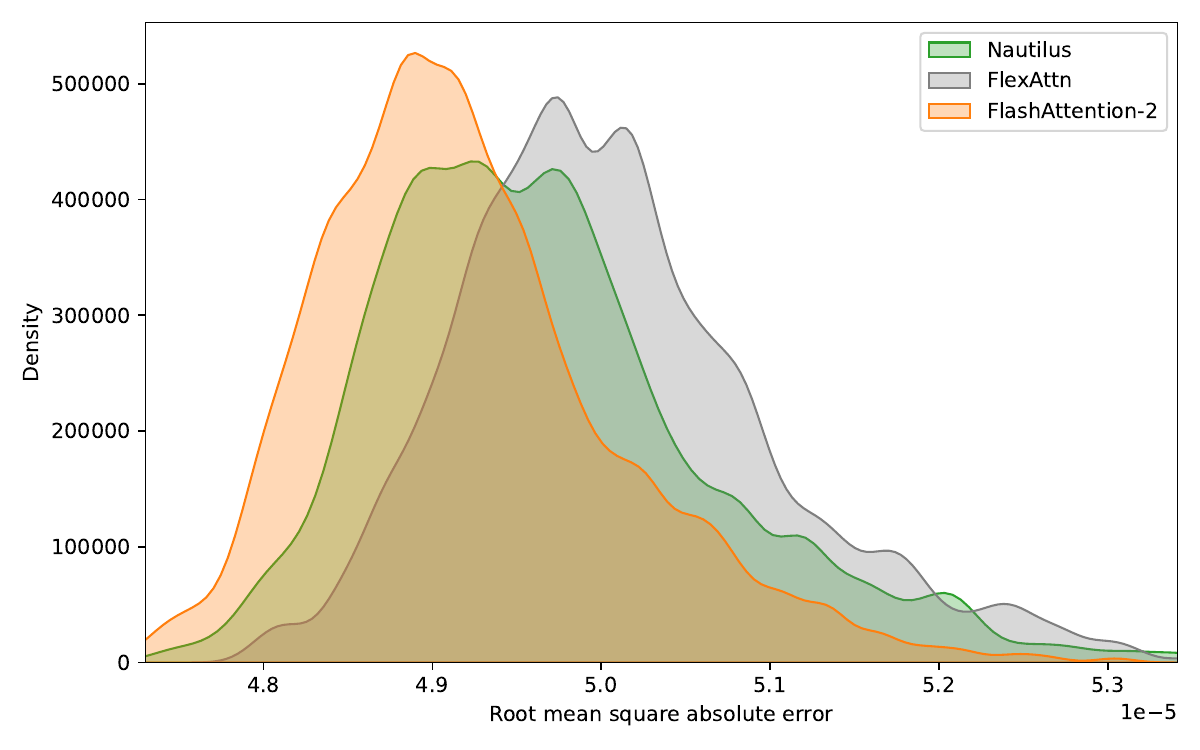}
    \caption{PF GQA.}
    \label{fig:rmse-kde-gqa}
  \end{subfigure}

  \caption{Kernel numerical stability comparison on Qwen across three attention settings.}
  \label{fig:rmse-kde-all}
\end{figure}

In Table \ref{tab:numeric_error_absolute_error}, we report the numerical error of the
three frameworks using root mean square error (RMS) together with the 90th and 99th
percentile error magnitudes. We also examine how per-sample (per-sentence) RMSE is
distributed. The results for PF Global, PF Causal, and PF GQA are shown in
Figure \ref{fig:rmse-kde-all}. Overall, Nautilus stays very close to Tri-Dao
Attention in error and is consistently better than FlexAttention.

\end{document}